# Stimulated emission from ZnO thin films with high optical gain and low loss

A-S. Gadallah[1,2], K. Nomenyo[1], C. Couteau[1,4], D. J. Rogers[3] and G. Lérondel[1]

[1]*Laboratoire de Nanotechnologie et d'Instrumentation Optique, Institut Charles Delaunay, CNRS UMR 6279, Université de Technologie de Troyes, 12 rue Marie Curie, BP 2060, 10010 Troyes Cedex, France*

[2]*Department of Laser Sciences and Interactions, National Institute of Laser Enhanced Sciences, Cairo University Giza, Egypt*

[3]*Nanovation, 8, route de Chevreuse, 78117 Châteaufort, France*

[4]*CINTRA, CNRS-NTU-Thales UMI 3288, Research Techno Plaza, 50 Nanyang Drive, Border X Block, Level 6, Singapore 637553, Singapore*

**Abstract:**

Stimulated surface- and edge-emission were investigated for ZnO thin films grown epitaxially by pulsed laser deposition. The lasing threshold was 0.32 MW/cm$^2$ for surface pumping and 0.5 MW/cm$^2$ for edge pumping, which is significantly lower than thresholds observed previously. A modified variable stripe length method was used to measure the gain, which was 1369 cm$^{-1}$ for N-band emission. Losses were measured using the shifting excitation spot method and values of 6.2 cm$^{-1}$ and 6.3 cm$^{-1}$ were found for the N-band and P-band, respectively. The measured gain and loss were the highest and lowest (respectively) ever reported for ZnO films.

There has been a huge resurge of interest in ZnO over the last 10 years [1]. This is due, in part, to appealing attributes such as a direct wide bandgap, relatively high exciton binding energy (~60 meV), comparatively low cost and good biocompatibility. These properties predispose ZnO for a variety of applications, including optoelectronics (e.g. light emitting diodes, laser diodes (LDs), solar cells and photodetectors), biomedical and sensing. Optical gain and optical loss are two key characteristics for opto-semiconductor thin films. Higher optical gain, for instance, allows amplification with shorter active-medium lengths, while reduced losses allow lower lasing thresholds. In the literature, stimulated emission has been observed from the surface [2, 3], from the edge [4,5] and even via random lasing [6,7] for ZnO thin films. Zhang et al. [4] reported the highest optical gain value for ZnO films, so far, of 571 $cm^{-1}$. This was for layers grown by laser molecular beam epitaxy [4]. Other work reported gains of 340 $cm^{-1}$ [8], 300 $cm^{-1}$ [5], 160 $cm^{-1}$ [2] and 40 $cm^{-1}$ [3]. Zhang et al. also reported the lowest optical losses for ZnO thin films, of 68 $cm^{-1}$ [4]. This paper investigates stimulated emission from both the surface and the edge of ZnO thin films grown epitaxially on c-sapphire substrates. Special emphasis was given to the study of optical gain and losses, which were found to be over 1000 $cm^{-1}$ and, below 10 $cm^{-1}$, respectively.

In this study, stimulated surface emission was pumped using an $N_2$ laser, emitting at 337 nm (with 5 ns pulse duration and 10 Hz repetition rate). The spot diameter was 1 mm. Stimulated edge emission was pumped with a frequency-tripled, Q-switched, Nd:YAG laser emitting at 355 nm (with 5 ns pulse duration and 10 Hz repetition rate). A laser stripe, 1 cm long and 200 μm wide, was focused on the sample surface using a cylindrical lens. The emission was collected from an end facet of the sample. In both cases, spectra were acquired using a standard photoluminescence set-up coupled to a spectrometer (50 cm focal length) with a Peltier-cooled charge coupled device camera.

For optical gain investigations, the variable stripe length (VSL) method was employed [9]. This involved controlling the length of a stripe excited by the pump laser with a cylindrical lens and recording the spectrum emitted from the edge of the film. For optical loss measurements, the shifting excitation spot (SES) method [10] was employed. This involved displacing the laser spot from the edge towards the middle of the sample and recording the position dependence of the photoluminescence spectrum. All measurements were carried out at room temperature.

ZnO thin films of various thicknesses, grown on a range of substrates, by different techniques (e.g. metal organic vapor deposition, pulsed laser deposition (PLD), chemical deposition) were measured. One layer, in particular, gave considerably more intense stimulated emission and the lowest lasing threshold. This film was grown by PLD on a c-sapphire substrate, using conditions described elsewhere [11]. The thickness was estimated to be about 500 nm by ellipsometry (in prior work [2], the authors reported that thinner films grown under similar conditions were found to exhibit random lasing [6,7]). High resolution x-ray diffraction (XRD) revealed the film to be epitaxial [12] with an omega rocking curve linewidth of 0.14° and a c lattice parameter of 5.209 Å, which is relatively close to the expected equilibrium value for wurtzite ZnO, of about 5.206 Å. Thus there was no indication of significant residual epitaxial strain at the film surface. Contact mode atomic force microscopy (AFM) revealed a root mean square roughness of about 2.5 nm with a peak-to-valley of 10 nm for a scan area of 2µm x 2µm. Average grain size was evaluated, from the XRD 2 theta-omega linewidth, to be around 300 nm. A slightly larger estimation was obtained for the grain size at the film surface by AFM. Four point resistivity measurements gave an average resistivity of about 0.3 Ω.cm, which is typical for as-grown ZnO films, which are slightly oxygen depleted. More detailed materials characterization data and analysis will be published elsewhere.

Surface photoluminescence spectra showed an intense ultraviolet (UV) near band edge (NBE) emission peak and very low signal in the green, which indicates that the defect density was relatively low. Figure 1-a presents surface emission spectra around the NBE peak as a function of N$_2$ laser excitation intensity. Below the lasing threshold, a single peak (centred at around 378 nm) characteristic of spontaneous emission was observed. This peak was attributed to radiative recombination of free excitons [13]. As the power density was increased to 0.32 MW/cm$^2$, an additional stimulated emission peak appeared at around 392 nm. This peak was attributed to a P-band exciton-exciton interaction. In such a process, one exciton is scattered to a higher energy state while the other emits a photon with an energy given by [14]:

$$E_n = E_{ex} - E_b^{ex}\left(1 - \frac{1}{n^2}\right) - \frac{3}{2}k_B T \qquad (1)$$

where E$_{ex}$ is the free exciton energy (eV), $E_b^{ex}$ is the exciton binding energy (eV) and n (=2, 3,…., ∞) is the quantum number of the excited exciton state. As the pump intensity was increased up to 0.51 MW/cm$^2$, another stimulated emission peak appeared at around 395 nm. This peak was attributed to an N-band electron-hole plasma emission (EHP) [15, 16]. In this case, the exciton concentration, n$_{ex}$ (number of excitons.cm$^{-3}$), is given by [17]:

$$n_{ex} = \frac{I_p \tau}{h\nu_p d} \qquad (2)$$

where $I_p$, $\tau$, $h\nu_p$, and $d$ are the pump intensity (MW/cm$^2$), exciton lifetime (taken to be 0.3 ns [18]), pump photon energy (J) and film thickness (cm), respectively. With these parameters, an I$_p$ of 0.51 MW/cm$^2$ gives an n$_{ex}$ of around 5.35 x 10$^{18}$ cm$^{-3}$. This concentration is higher than the Mott concentration for ZnO (about 10$^{17}$ cm$^{-3}$ [19]), which implies that charge screening causes excitons to lose their individual nature and form an EHP. By further increasing the pump power density, the N-band intensity exceeded that of the P-band, which is indicative of a tendency towards EHP emission. This interpretation was corroborated by a red-shift in the N-band, which was attributed to band-gap renormalization [20]. Figure 1-b presents the integrated emission intensity as a function of pump power. Below the lasing threshold, spontaneous emission occurred by free exciton recombination, which presented a linear dependence on pump power density. Above this threshold, the integrated emission intensity dependence on pump power density showed a much steeper slope and peak narrowing from about 22 nm full width at half maximum (FWHM) below threshold to about 5.5 nm FWHM above threshold. The integrated emission showed a power dependence (m = 2.5) on the pump power density. Such non-linear dependence combined with peak narrowing are typical signatures of the onset of stimulated emission.

A similar trend was found for edge emission. Figure 2 presents the emission spectra collected from one end facet for different pump intensities. For a pump intensity of 0.25 MW/cm$^2$, the P-band emerged at 391 nm. Then, at intensities above the lasing threshold (0.5 MW/cm$^2$), the N-band stimulated emission appeared at 394 nm. Interestingly, the threshold intensity increased compared to the surface emission configuration (0.32 MW/cm$^2$). This was attributed to the different geometry of the pumped area [2,4]. Such a change in geometry modifies the confinement factor, $\Gamma$, and hence the threshold intensity (which is proportional to exp $(1/\Gamma)$ [21]). As for the surface emission, a red-shift was also seen for the edge stimulated emission. The output stimulated emission showed an m = 2.26 power dependence on excitation intensity. Peak narrowing from about 18 nm FWHM below threshold to about 9 nm FWHM above threshold was also observed. For films that do not show stimulated emissions process, neither the peak narrowing in the spectrum nor the two distinct slopes in the integrated output intensity versus pump intensity were observed. A lasing-mode photo of the sample for above-threshold pumping is shown in the inset of Fig. 2-b. For the gain VSL method, the amplified spontaneous emission (ASE) intensity $I_{ASE}(\lambda,L)$ (W.cm$^{-2}$) is related to the gain $g(\lambda)$ through [9]

$$I_{ASE} = \frac{A}{g(\lambda)}[\exp(g(\lambda)L)-1] \qquad (3)$$

where A (W.cm$^{-1}$) and L (cm) are the scaling factor and the exited stripe length, respectively. In this model, g is assumed to be independent of L in the unsaturated region. Hence, for evaluating g, L is changed and the corresponding $I_{ASE}$ ($\lambda$,L) is recorded. By fitting the curve of $I_{ASE}$ ($\lambda$,L) versus L, Fig. 3-a, with the above equation in the unsaturated region, the gain is obtained [5, 22]. Figure 3-a shows the recorded $I_{ASE}$($\lambda$, L) versus L at 391 nm and 394 nm (corresponding to the P-band and N-band, respectively). The gain was measured to be 861 cm$^{-1}$ at 391 nm and 956 cm$^{-1}$ at 394 nm. To the best of our knowledge, these are the highest gains ever reported for ZnO thin films. Two possible explanations are put forward for this high gain in our case. The first is a probable reduction in dislocation density compared with the thinner films measured previously [2] because of a tendency for dislocation density to diminish during growth of such oxygen-depleted layers [23]. Lower defect density reduces exciton quenching, and thereby increases the gain. Transmission electron microscopy will be pursued in order to investigate this hypothesis. The second suggested explanation is a result of weak exciton confinement due to the relatively large grain size. This is based on work suggesting that grain boundaries confine the excitons, which leads to increased $n_{ex}$ [24] which augments the gain.

According to the modified VSL method, the gain depends not only on the wavelength but also on the excitation length, L, and is given by [8]

$$g(\lambda,L) = \frac{\left(dI_{ASE}(\lambda,L)/dL\right) - A}{I_{ASE}(\lambda,L)} \qquad (4)$$

where $dI_{ASE}(\lambda,L)/dL$ is the first derivative of $I_{ASE}(\lambda,L)$ with respect to $L$. The constant $A$ is obtained from $dI_{ASE}(\lambda,L)/dL$ at $L = 0$. Figure 3-b shows the gain as a function of $L$ at 391 nm (P-band) and 394 nm (N-band) using the modified VSL, presented above. The gain is maximum at an $L$ of 60 μm for both wavelengths: 758 cm$^{-1}$ at 391 nm and 1369 cm$^{-1}$ at 394 nm. Interestingly, in Figure 3-b there is oscillation in the measured gain. Such oscillations have been elsewhere and, according to Kim et al, they might be due photon-carrier interaction that leads to a coupled mode along the stripe [8]. The origin of this oscillation remains, however, an open question. A common finding for the two models (standard and modified VSL) is that the gain for the N-band is higher than that for the P-band.

For measuring the SES optical losses, the pump intensity was constant during measurement and equal to 1.3 MW/cm$^2$ (higher than the lasing threshold). The emission intensity as a function of the distance from the sample edge for wavelengths of 391 nm and 394 nm is shown in Fig. 4-a. By fitting these curves, the losses at 391 nm and 394 nm were found to be 6.3 cm$^{-1}$ and 6.2 cm$^{-1}$, respectively. The losses here are believed to be mainly due to interface scattering, as confirmed by AFM images (not shown). To the best of our knowledge, these are the lowest losses measured so far for ZnO films (Zhang et al. [4] previously reported optical losses of 68 cm$^{-1}$ for ZnO thin films). However, they are still higher than that of GaN films grown by molecular beam epitaxy [26], which can be as low as 1.23 cm$^{-1}$.

Detailed analysis of surface- and edge-emitting stimulated emission using optical pumping has been presented. Typical lasing thresholds of 0.32 MW/cm$^2$ for surface pumping and 0.5 MW/cm$^2$ for edge-emitting geometry were measured. These values were relatively low compared to those reported in previous work [2] but not remarkable. The measured gain and loss, however, were the highest and lowest (respectively) ever reported for ZnO films. Comparatively low surface defect density, due to dislocation die-out during film growth, and weak exciton confinement, due to a relatively large grain size, were put forward as possible origins. These results confirm the exceptional optical quality of these ZnO thin films grown on sapphire by PLD and their potential for use in next-generation UV photonic devices.

This work was supported by the CPER MATISSE project. Three of the authors, A. G, K. N. , and D. J. R. would like to thank respectively Campus France, the "region Champagnes-Ardennes", the European Social Fund for financial support and Centrale de Technologie Universitaire IEF-MINERVE at Orsay University for access to the X-Ray diffraction equipment.

**Figures captions**

Fig.1 a) Surface emission spectra for different pumping intensities, b) integrated emission intensity versus pump intensity for the ZnO thin film. Scattered squares are the measured values and the intersection point between the two best-fit lines is the lasing threshold (0.32 MW/cm$^2$).

Fig.2. a) Edge emission spectra for different pumping intensities, b) integrated emission intensity versus pump intensity for the ZnO thin film. Scattered squares are the measured values and the intersection between the two best-fit lines is the lasing threshold (0.5 MW/cm$^2$). Inset is a lasing-mode photo for the sample (1cm x 0.5cm).

Fig.3: Emission intensity (a) and gain (b) versus excitation length at 391 nm and 394 nm.

Fig.4: Emission intensity versus distance from the sample edge at 391 nm and 394 nm.

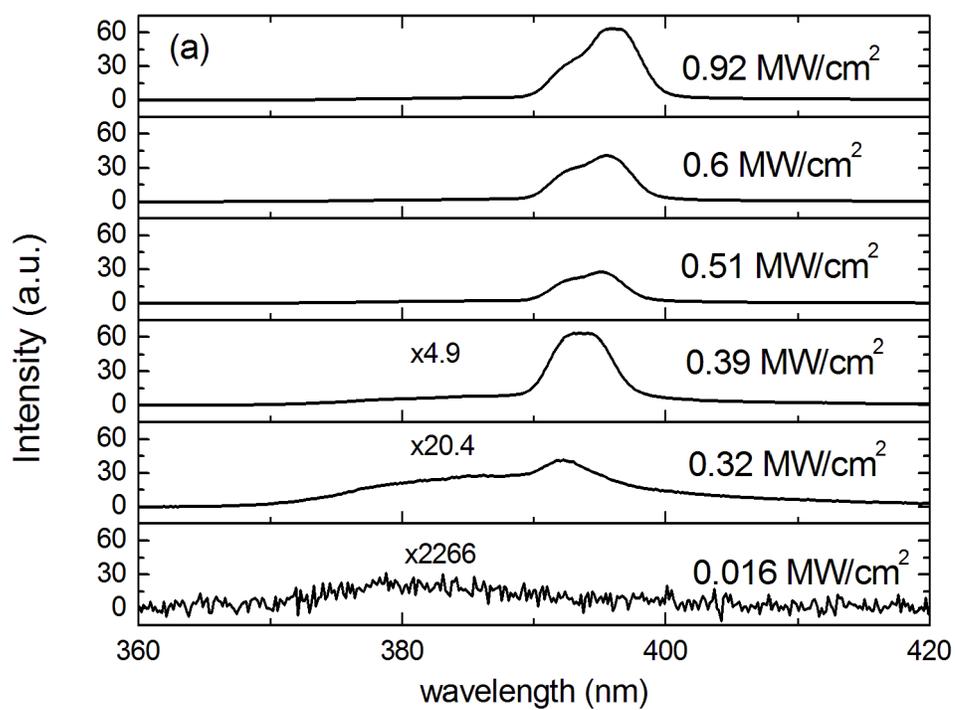

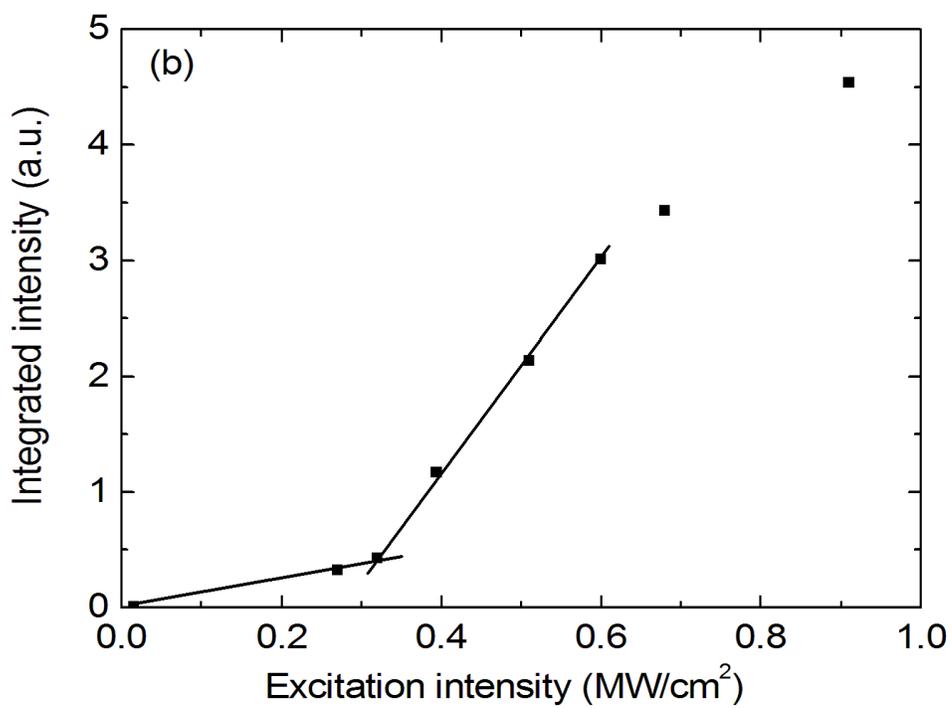

Fig.1

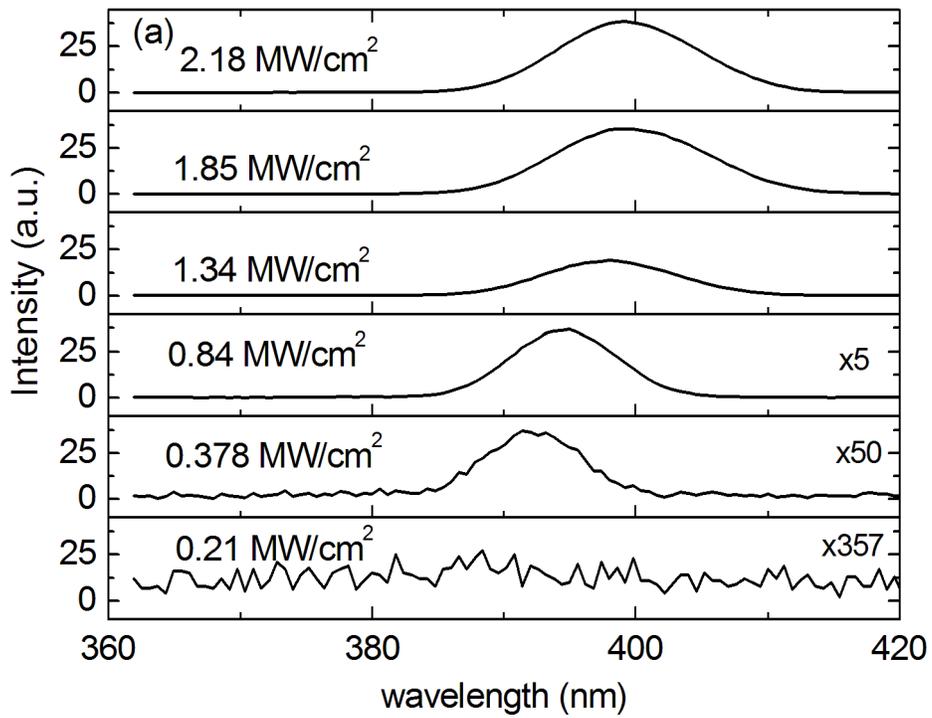
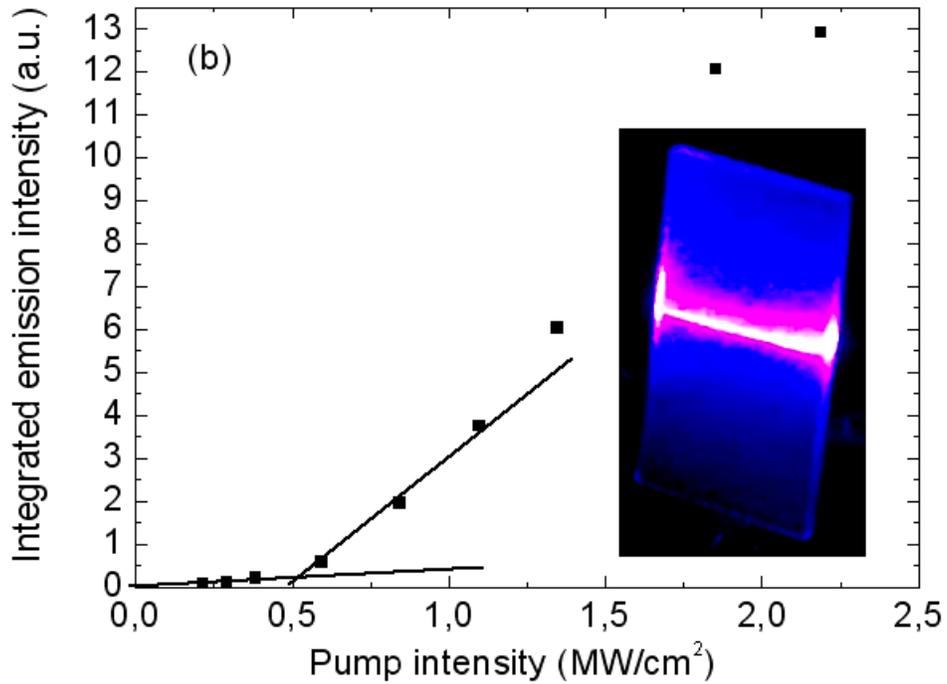

Fig.2.

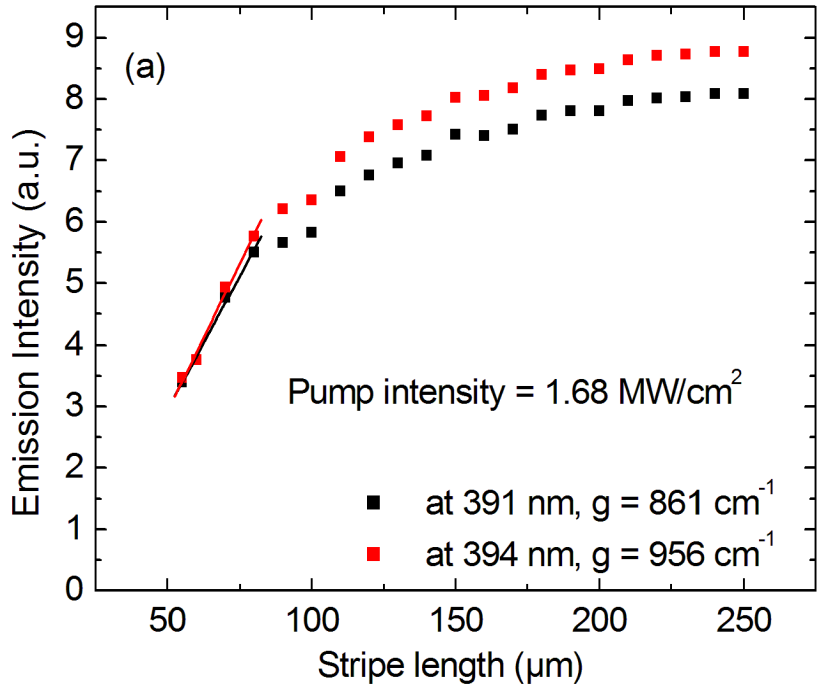
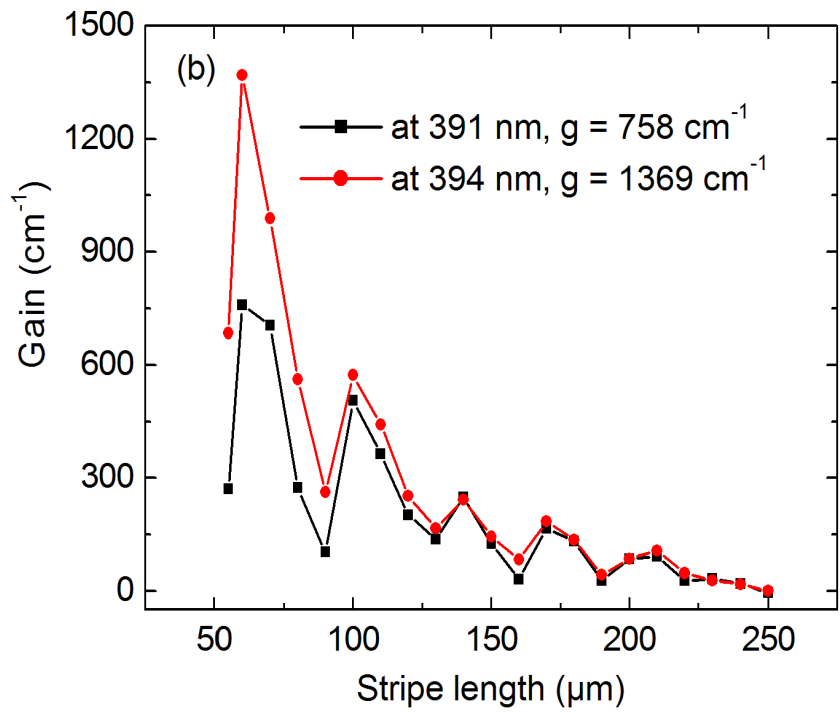

Fig.3

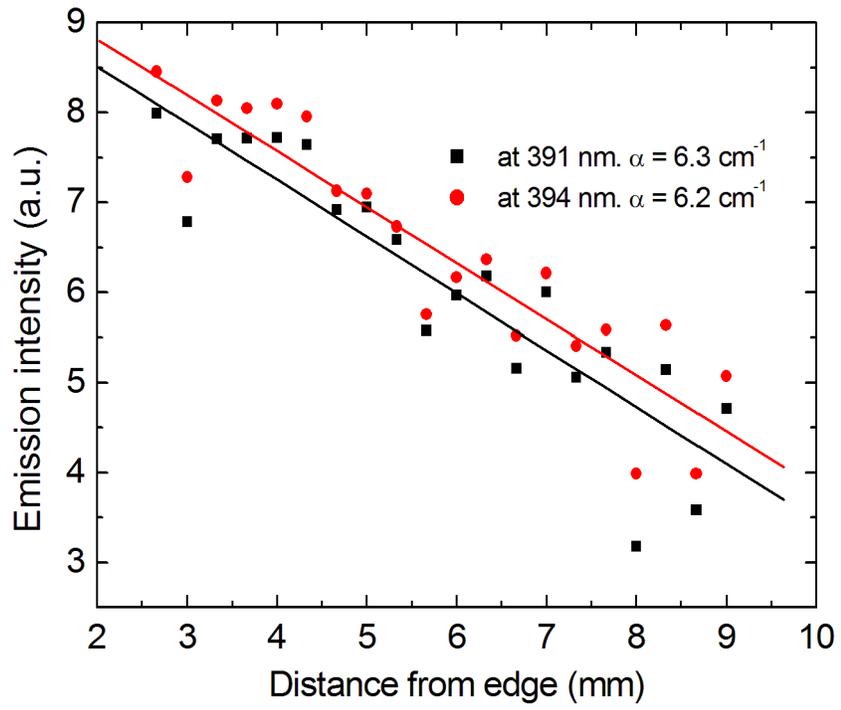

Fig.4